\documentclass[aps,prd,onecolumn,showkeys,amssymb]{revtex4}
\usepackage{float}
\usepackage{amssymb}
\usepackage{graphicx}
\usepackage{amsmath}
\usepackage{hyperref}
\begin{document}

\title{Further stable neutron star models from $f(R)$ gravity}

\author{Artyom V. Astashenok$^{1}$, Salvatore Capozziello$^{2,3}$, Sergei D. Odintsov$^{4,5,6}$}

\affiliation{$^{1}$I. Kant Baltic Federal University, Institute of Physics and Technology, Nevskogo st. 14, 236041 Kaliningrad, Russia,\\
$^2$Dipartimento di Fisica, Universita' di Napoli "Federico II
and \\$^3$INFN Sez. di Napoli, Compl. Univ. di Monte S. Angelo, Ed. G., Via Cinthia,
9, I-80126, Napoli, Italy,\\
$^4$Instituci\`{o} Catalana de Recerca i Estudis Avan\c{c}ats (ICREA), Barcelona, Spain\\
$^5$Institut de Ciencies de l'Espai (IEEC-CSIC), Campus UAB, Facultat de Ciencies, Torre C5-Par-2a pl, E-08193 Bellaterra, Barcelona, Spain\\
$^6$Tomsk State Pedagogical University (TSPU), Tomsk,
Russia }

\begin{abstract}
Neutron star models in perturbative $f(R)$ gravity are considered  with
realistic equations of state. In particular, we consider the FPS, SLy and other equations of state and a case of piecewise equation of state for stars with quark cores. The mass-radius relations for   $f(R)=R+R(e^{-R/R_{0}}-1)$ model and for  $R^2$ models with logarithmic and cubic corrections are obtained.   In the case of $R^2$ gravity with cubic corrections, we obtain that at high central densities ($\rho>10\rho_{ns}$, where $\rho_{ns}=2.7\times 10^{14}$ g/cm$^{3}$ is the nuclear saturation density),  stable star configurations  exist. The minimal radius of such stars is close to $9$ km with maximal mass $\sim 1.9 M_{\odot}$ (SLy equation). A similar situation takes place for AP4 and BSK20 EoS. Such an effect can give rise to more compact stars than in General Relativity. If observationally identified, such objects could constitute a formidable signature for modified gravity at astrophysical level. Another interesting result can be achieved in modified gravity with only a cubic correction. For some EoS, the upper limit of neutron star mass increases and therefore these EoS can describe realistic star configurations (although, in General Relativity, these EoS are excluded by observational constraints).
\end{abstract}


\maketitle

\section{Introduction}

The current accelerated expansion of the universe has been confirmed
by several  independent observations. Standard candles and  distance indicators point out an accelerated expansion which cannot be obtained by ordinary perfect fluid matter as source for the cosmological Friedmann equations \cite{Perlmutter, Riess1, Riess2}. This evidence  gives rise to  difficulties  in order to explain the  evolution of  large scale structures. Furthermore observations  of microwave background radiation (CMBR) anisotropies \cite{Spergel},  of cosmic shear through gravitational weak leasing surveys \cite{Schmidt} and, finally, data on Lyman alpha forest absorption lines \cite{McDonald} confirm the picture of an accelerated Hubble fluid.	

In particular, the  discrepancy between the amount of  luminous matter revealed from observations  and the critical density needed to obtain a spatially flat universe could be solved if one assumes the existence of a non-standard cosmic fluid with negative pressure, which is not clustered in large scale structure. In the simplest scenario, this {\it dark energy}, can be addressed as the Einstein Cosmological Constant and would contribute about 70\% to the global energy budget of the universe. The remaining 30\%,  clustered in galaxies and clusters of galaxies, should be constituted for about 4\% by baryons and for the rest by cold dark matter (CDM), whose candidates, at fundamental level, could be WIMPs (Weak Interacting Massive Particles), axions or other unknown particles \cite{Krauss}.

From an observational viewpoint, this  model has the  feature to be in  agreement with data coming from observations. It could be  assumed as the first step towards a new standard cosmological model and it is  indicated as Concordance Lambda Cold Dark Matter ($\Lambda$CDM) Model \cite{Bancall}. In summary, the  observed universe could be self-consistently described once we admit the presence of a cosmological constant (70\% of the total content), which would give rise to the observed acceleration of the Hubble fluid, and the presence of dark matter (at least 25\%), which would explain the large scale structure. Despite of the  agreement with observations, the $\Lambda$CDM model presents  incongruences from a theoretical viewpoint. If the cosmological constant constitutes the ``vacuum state'' of the gravitational field, we have to explain the 120 orders of magnitude between its  observed value at cosmological level and the one predicted by any quantum gravity \cite{Weinberg}. This inconsistency, also known as the  {\it cosmological constant problem}, is one of the most fundamental problems of cosmology.

A very straightforward approach is to look for explanations for dark matter and dark energy within the realm of known physics. On the other hand, an  alternative is that General Relativity is not capable  of describing  the universe at scales larger than Solar System, and  dark components (energy + matter) could be the observable effect of such an inadequacy.

Assuming this point of view,  one can propose alternative theories of gravity extending the Einstein theory (in this sense one deals with  modified  gravity), keeping its positive results, without requiring  dark components,  up to now not detected at experimental level. In this perspective, it can be shown that the accelerated expansion can be obtained without using new fundamental ingredients but enlarging the gravitational sector (see for example  \cite{Capozziello1, Capozziello2, Odintsov1, Turner, Odintsov-3, Capozziello3,Capozziello_book, Capozziello4, Cruz}).

In particular, it has been recently shown that such theories give models able to reproduce the Hubble diagram derived from SNela surveys \cite{Capozziello3,Demianski} and the anisotropies observed for CMBR \cite{Perrotta, Hwang}.

However, also this approach needs  new signatures or  some {\it experimentum crucis} in order to be accepted or refuted. In particular, exotic astrophysical structures, which cannot be addressed by standard gravity,  could constitute a powerful tool to address this problem. In particular, strong field regimes of relativistic astrophysical objects could discriminate between General Relativity and its possible extensions.

The study of relativistic stars in modified gravity could have very interesting consequences to address this issue. In fact, new  theoretical stellar structures emerge and they  could have very important observational consequences constituting the signature  for the Extended Gravity (see e.g. \cite{Laurentis, Laurentis2}).  Furthermore,  strong gravitational regimes could be considered if one assume General Relativity as the weak field limit of some more complicated effective gravitational theory \cite{Dimitri-rev}.  In particular, considering the simplest extension of General Relativity, namely the $f(R)$ gravity, some models can be rejected because  do not allow the existence of stable star configurations  \cite{Briscese, Abdalla, Bamba, Kobayashi-Maeda, Nojiri5}. On the other hand, stability can be achieved in certain cases due to the so called  {\it Chameleon Mechanism} \cite{Tsujikawa, Upadhye-Hu}.  Another problem is that the possibility of existence of stable star configuration may depend on the  choice of equation of state (EoS). For example, in  \cite{Babichev1, Babichev2},  a polytropic EoS is used in order to solve  this issue although the adopted  EoS does not  seem  realistic to achieve reliable neutron stars.

In this paper, we start from the fact that $f(R)$ gravity models  introduce a new scalar degree of freedom that must be considered into dynamics, then  we study the structure of neutron stars in perturbative $f(R)$ gravity  where the scalar curvature $R$ is defined by Einstein equations at zeroth order on the small parameter, i.e. $R\sim T$, where $T$ is the trace of energy-momentum tensor.

In this framework, we investigate  several  $f(R)$ models,  namely  $f(R)=R+\beta R (\exp(-R/R_{0})-1)$,  $R^2$ model with logarithmic [$f(R)=R+\alpha R^{2} (1+\beta ln (R/\mu^{2})$] and cubic [$f(R)=R+\alpha R^{2}(1+\gamma R)$] corrections. {In particular, we consider the FPS and SLy equations of state for exponential modified gravity and a case of piecewise EoS for neutron stars with quark cores for logarithmic model. For models with a cubic term correction,  the cases of realistic EoS such as SLy, AP4 and BSK20 are considered.} One of the  results is that, if cubic correction term,  at some   densities, is comparable with the quadratic one,   stable star configurations  exist at high central densities.  The minimal radius of such stars is close to $9$ km for maximal mass $\sim 1.9 M_{\odot}$ (SLy equation) or to $8.5$ km for mass $\sim 1.7M_{\odot}$ (FPS equation). {It is interesting to note that, in the case of simple cubic gravity,  the maximal mass of stable configurations may be greater than the maximal limit of mass in the case of General Relativity. Due to this effect,  some EoS, which are ruled out by observational constraints in GR, can lead to realistic results in the context  of modified gravity. Moreover, these EoS describe some observational data with better precision than in General Relativity.} Clearly, such objects cannot be achieved in the context of General Relativity \cite{werner} so their possible observational evidences could constitute a powerful probe for modified gravity \cite{felicia,antoniadis,freire}.

The  paper is organized as   follows. In Section II, we present the field equations for $f(R)$ gravity. For spherically symmetric solutions of these equations, we obtain the modified
Tolman--Oppenheimer--Volkoff (TOV) equations. These equations are  numerically solved   by a  perturbative approach considering  realistic equations of state in Section III. In this context, new stable structures, not existing in General Relativity, clearly emerge. Discussion of the results and conclusions  are reported in Sec. IV.

\section{Modified TOV equations in $f(R)$ gravity}

Let us  start from the action for $f(R)$ gravity. Here the  Hilbert-Einstein action, linear in the Ricci curvature scalar $R$,  is  replaced by a generic  function $f(R)$:
\begin{equation}\label{action}
S=\frac{c^4}{16\pi G}\int d^4x \sqrt{-g}f(R) + S_{{\rm matter}}\quad ,
\end{equation}
where $g$ is determinant of the metric $g_{\mu\nu}$ and $S_{\rm matter}$ is the action of the standard perfect fluid matter.
The field equations  for metric $g_{\mu\nu}$ can be obtained by varying with respect to  $g_{\mu\nu}$.  It is convenient to write function $f(R)$ as
\begin{equation}\label{fR}
    f(R)=R+\alpha h(R),
\end{equation}
where
$h(R)$ is, for now, an arbitrary function. In this notation,  the field equations are
\begin{equation}\label{field}
(1+\alpha h_{R})G_{\mu \nu }-\frac{1}{2}\alpha(h-h_{R}R)g_{\mu \nu
}-\alpha (\nabla _{\mu }\nabla _{\nu }-g_{\mu \nu }\Box )h_{R}=\frac{8\pi
G}{c^4} T_{\mu \nu }\,.
\end{equation}
Here $G_{\mu\nu}=R_{\mu\nu}-\frac12Rg_{\mu\nu}$ is the Einstein
tensor and ${\displaystyle h_R=\frac{dh}{dR}}$ is the derivative of $h(R)$ with
respect to the scalar curvature.
We are searching for  the solutions of these equations assuming a spherically
symmetric metric with two independent functions of radial coordinate, that is:
\begin{equation}\label{metric}
    ds^2= -e^{2\phi}c^2 dt^2 +e^{2\lambda}dr^2 +r^2 (d\theta^2
    +\sin^2\theta d\phi^2).
\end{equation}
The energy--momentum tensor in the r.h.s. of Eq. (\ref{field}) is that
 of a perfect fluid, i.e. $T_{\mu\nu}=\mbox{diag}(e^{2\phi}\rho c^{2}, e^{2\lambda}P, r^2P, r^{2}\sin^{2}\theta P)$, where $\rho$ is the matter density and $P$ is the pressure. The components of the  field equations can be written as
\begin{eqnarray}
 \frac{ -8\pi G}{c^2} \rho &=& -r^{-2} +e^{-2\lambda}(1-2r\lambda')r^{-2}
                +\alpha h_R(-r^{-2} +e^{-2\lambda}(1-2r\lambda')r^{-2}) \nonumber \\
             && -\frac12\alpha(h-h_{R}R) +e^{-2\lambda}\alpha[h_R'r^{-1}(2-r\lambda')+h_R''] \label{f-tt},\\
  \frac{8\pi G}{c^4} P &=& -r^{-2} +e^{-2\lambda}(1+2r\phi')r^{-2}
                +\alpha h_R(-r^{-2} +e^{-2\lambda}(1+2r\phi')r^{-2}) \nonumber \\
             && -\frac12\alpha(h-h_{R}R) +e^{-2\lambda}\alpha h_R'r^{-1}(2+r\phi'), \label{f-rr}
\end{eqnarray}
where prime denotes derivative with respect to radial distance, $r$.
For the exterior solution,  we assume a Schwarzschild solution. For this reason, it is convenient to define the change of variable  \cite{Stephani,Cooney}
\begin{equation}\label{mass}
    e^{-2\lambda}=1-\frac{2G M}{c^2 r}\,.
\end{equation}
The value of parameter $M$ on the surface of a neutron stars can be considered as a gravitational star mass. The following  relation
\begin{equation}\label{dMa/dr}
    \frac{G dM}{c^{2}dr}=\frac{1}{2}\left[1-e^{-2\lambda}(1-2r\lambda']\right)\,,
\end{equation}
is useful for the derivative $dM/dr$.

The  hydrostatic condition equilibrium can be obtained from the Bianchi identities which give  conservation equation of the  energy-momentum tensor,   $\nabla^\mu T_{\mu\nu}=0$, that, for a perfect fluid,  is
\begin{equation}\label{hydro}
    \frac{dP}{dr}=-(\rho
    +P/c^2)\frac{d\phi}{dr}\,,.
\end{equation}
The second TOV equation can be obtained by substitution of the derivative $d\phi/dr$ from (\ref{hydro}) in Eq.(\ref{f-rr}). Then we  use the dimensionless variables defined according to the substitutions
$$
M=m M_{\odot},\quad r\rightarrow r_{g}r, \quad \rho\rightarrow\rho M_{\odot}/r_{g}^{3},\quad P\rightarrow p M_{\odot}c^{2}/r_{g}^{3}, \quad R\rightarrow {R}/r_{g}^{2}.
$$
Here $M_{\odot}$ is the Sun mass and $r_{g}=GM_{\odot}/c^{2}=1.47473$ km. In terms of these variables,  Eqs. (\ref{f-tt}), (\ref{f-rr}) can be rewritten, after some manipulations, as
\begin{equation}\label{TOV-1}
\left(1+\alpha r_{g}^{2} h_{{R}}+\frac{1}{2}\alpha r_{g}^{2} h'_{{R}} r\right)\frac{dm}{dr}=4\pi{\rho}r^{2}-\frac{1}{4}\alpha r^2 r_{g}^{2}\left(h-h_{{R}}{R}-2\left(1-\frac{2m}{r}\right)\left(\frac{2h'_{{R}}}{r}+h''_{{R}}\right)\right),
\end{equation}
\begin{equation}\label{TOV-2}
8\pi p=-2\left(1+\alpha r_{g}^{2}h_{{R}}\right)\frac{m}{r^{3}}-\left(1-\frac{2m}{r}\right)\left(\frac{2}{r}(1+\alpha r_{g}^{2} h_{{R}})+\alpha r_{g}^{2} h'_{{R}}\right)({\rho}+p)^{-1}\frac{dp}{dr}-
\end{equation}
$$
-\frac{1}{2}\alpha r_{g}^{2}\left(h-h_{{R}}{R}-4\left(1-\frac{2m}{r}\right)\frac{h'_{{R}}}{r}\right),
$$
where $'=d/dr$. For $\alpha=0$, Eqs. (\ref{TOV-1}), (\ref{TOV-2})  reduce to
\begin{equation}
\frac{dm}{dr}=4\pi\tilde{\rho} r^{2}
\end{equation}
\begin{equation}
\frac{dp}{dr}=-\frac{4\pi p r^{3}+m}{r(r-2m)}\left(\tilde{\rho}+p\right),
\end{equation}
i.e. to ordinary dimensionless TOV equations. These equations can be solved numerically for a given EoS $p=f({\rho})$ and initial conditions $m(0)=0$ and ${\rho}(0)={\rho}_{c}$.

For non-zero $\alpha$, one needs the third equation for the Ricci  curvature scalar. The trace of  field Eqs. (\ref{field}) gives the  relation
\begin{equation}
3\alpha\square h_{R}+\alpha h_{R}R-2\alpha h-R=-\frac{8\pi G}{c^{4}}(-3P+\rho c^{2}).
\end{equation}
In dimensionless variables, we have
\begin{equation}\label{TOV-3}
3\alpha r_{g}^{2}\left(\left(\frac{2}{r}-\frac{3m}{r^{2}}-\frac{dm}{rdr}-\left(1-\frac{2m}{r}\right)\frac{dp}{(\rho+p)dr}\right)\frac{d}{dr}+
\left(1-\frac{2m}{r}\right)\frac{d^{2}}{dr^{2}}\right)h_{{R}}+\alpha r_{g}^{2} h_{{R}}{R}-2\alpha r_{g}^{2} h-{R}=-8\pi({\rho}-3p)\,.
\end{equation}
One has to note that the combination $\alpha r_{g}^{2} h(R)$ is a dimensionless function. We need to add the EoS for matter inside star to the Eqs. (\ref{TOV-1}), (\ref{TOV-2}), (\ref{TOV-3}). For the sake of simplicity, one can use the polytropic EoS $p\sim \rho^{\gamma}$ although a more realistic EoS  has to take into account  different physical states for different regions of the
star and it  is more complicated. With these considerations in mind, let us face the problem to construct neutron star models in the context of $f(R)$ gravity.

\section{Neutron star models in $f(R)$ gravity}

The solution of Eqs. (\ref{TOV-1})-(\ref{TOV-3}) can be achieved by using a perturbative approach (see \cite{Arapoglu,Alavirad} for details). For a perturbative solution the density, pressure, mass and curvature can be  expanded as
\begin{equation}
p=p^{(0)}+\alpha p^{(1)}+...,\quad \rho=\rho^{(0)}+\alpha \rho^{(1)}+...,
\end{equation}
$$
m=m^{(0)}+\alpha m^{(1)}+...,\quad R=R^{(0)}+\alpha R^{(1)}+...,
$$
where functions $\rho^{(0)}$, $p^{(0)}$, $m^{(0)}$ and $R^{(0)}$  satisfy to standard TOV equations assumed at zeroth order. Terms containing
$h_{R}$  are assumed  to be of first order in the small
parameter $\alpha$, so all such terms should be evaluated at   ${\mathcal O}(\alpha)$ order.
We have, for the $m=m^{(0)}+\alpha m^{(1)}$, the following equation
\begin{equation}
\frac{dm}{dr}=4\pi\rho r^2-\alpha r^{2}\left(4\pi \rho^{(0)}h_{R}+\frac{1}{4}\left(h-h_{R}R\right)\right)+\frac{1}{2}\alpha\left(\left(2r-3m^{(0)}-4\pi\rho^{(0)}r^{3}\right)\frac{d}{dr}+r(r-2m^{(0)})\frac{d^{2}}{dr^{2}}\right)
h_{R}
\end{equation}
and for pressure $p=p^{(0)}+\alpha p^{(1)}$
\begin{equation}
\frac{r-2m}{\rho+p}\frac{dp}{dr}=4\pi r^2 p+\frac{m}{r}-\alpha r^2\left(4\pi p^{(0)}h_{R}+\frac{1}{4}\left(h-h_{R}R\right)\right)-
\alpha \left(r-3m^{(0)}+2\pi p^{(0)}r^{3}\right)\frac{dh_{R}}{dr}.
\end{equation}
The Ricci curvature scalar,  in terms containing $h_{R}$ and $h$, has to be evaluated at  ${\mathcal O}(1)$ order, i.e.
\begin{equation}
R \thickapprox R^{(0)}=8\pi(\rho^{(0)}-3p^{(0)})\,.
\end{equation}
In this perturbative approach, we do not consider the  curvature scalar  as an additional degree of freedom since  its value is fixed by this relation.

{We can consider various EoS for the description of the behavior of nuclear matter at high densities.} {It is convenient to use analytical representations of these EoS. For example the SLy \cite{SLy} and FPS \cite{FPS} equation have the same analytical representation:}
\begin{equation}\label{FPS}
\zeta=\frac{a_{1}+a_{2}\xi+a_{3}\xi^3}{1+a_{4}\xi}f(a_{5}(\xi-a_{6}))+(a_{7}+a_{8}\xi)f(a_{9}(a_{10}-\xi))+
\end{equation}
$$
+(a_{11}+a_{12}\xi)f(a_{13}(a_{14}-\xi))+(a_{15}+a_{16}\xi)f(a_{17}(a_{18}-\xi)),
$$
where
$$
\zeta=\log(P/\mbox{dyn} \mbox{cm}^{-2})\,, \qquad \xi=\log(\rho/\mbox{g}\mbox{cm}^{-3})\,, \qquad f(x)=\frac{1}{\exp(x)+1}\,.
$$
The coefficients $a_{i}$ for SLy and FPS EoS are given in \cite{Camenzind}. {In \cite{Potekhin} the parametrization for three EoS models (BSK19, BSK20, BSK21) is offered. Also analytical parametrization for wide range of various EoS (23 equations) can be found in \cite{Eksi}.}

Furthermore,  we can consider the model of neutron star with a quark core. The quark matter can be described by the very simple EoS:
\begin{equation}\label{EoSQM}
p_{Q}=a(\rho-4B),
\end{equation}
where $a$ is a constant and the  parameter $B$ can vary from $\sim 60$ to $90$ Mev/fm$^{3}$. For quark matter with massless strange quark, it is  $a=1/3$. We consider  $a=0.28$  corresponding  to $m_{s}=250$ Mev. For numerical calculations,  Eq. (\ref{EoSQM}) is used for $\rho \geq \rho_{tr}$, where $\rho_{tr}$  is  the transition density {for which the pressure of quark matter coincides with the pressure of ordinary dense matter. } For example for FPS equation, the  transition density is $\rho_{tr}=1.069\times 10^{15}$ g/cm$^{3}$ ($B=80$ Mev/fm$^{3}$), for SLy equation $\rho_{tr}=1.029\times 10^{15}$ g/cm$^{3}$ ($B=60$ Mev/fm$^{3}$).
These parameters allow to set up neutron star models according to given $f(R)$ gravity models.

\textbf{Model 1}. Let's consider the simple exponential model
\begin{equation}\label{EXP}
f(R)=R+\beta R(\exp(-R/R_{0})-1),
\end{equation}
where $R_{0}$ is a constant. Similar models are considered in cosmology, see for example  \cite{EXP}. We can assume, for example, $R=0.5 r_{g}^{-2}$. For $R<<R_{0}$ this model  coincides with quadratic model of $f(R)$ gravity. The neutron stars models in frames of quadratic gravity is investigated in detail in \cite{Arapoglu}. It is interesting to consider the model (\ref{EXP}) for the investigations of higher order effects.

For neutron stars models with quark core,  one can see that there is no significant differences with respect to General Relativity. For a given central density,  the star mass grows with $\beta $. The dependence is close to linear for $\rho\sim 10^{15} \mbox{g/cm}^{3}$. For the piecewise equation of state (we consider the FPS case for $\rho<\rho_{tr}$) the maximal mass grows with increasing $\beta$. For $\beta=-0.25$, the maximal mass is $1.53M_{\odot}$, for $\beta=0.25$, $M_{max}=1.59M_{\odot}$ (in General Relativity, it is  $M_{max}=1.55M_{\odot}$). With an increasing $\beta$,  the maximal mass is reached at lower central densities. Furthermore, for  $dM/d\rho_{c}<0$,  there are no stable star configurations. A similar situation  is observed in the SLy case but mass grows with $\beta$ more slowly. It is interesting to stress that the $\beta$ parameter affects also the Jeans instability of any process that from self-gravitating systems leads to stellar formation as reported in \cite{Laurentis}.

For the simplified EoS (\ref{FPS}), other  interesting effects can occur. For  $\beta\sim -0.15$ at high central densities ($\rho_{c}\sim 3.0 - 3.5\times 10^{15}\mbox{g/cm}^{3}$), we have the dependence of the neutron star mass from radius (Figs. 1, 3) and from central density central density (see Figs. 2, 4). For $\beta<0$ for high central densities we have the stable star configurations ($dM/d\rho_{c}>0$).

{Of course the model with FPS EOS is ruled out by recent observations \cite{antoniadis, Demorest}. For example the measurement of mass of the neutron star PSR J1614-2230 with $1.97\pm0.04$ $M_{\odot}$ provides a stringent constraint on any $M-R$ relation. The model with SLy equation is more interesting: in the context of model (\ref{EXP}),  the upper limit of neutron star mass is around $2M_{\odot}$ and there is second branch of stability star configurations at high central densities. This branch describes observational data better than the model with SLy EoS in GR (Fig. 3).}

Although the applicability of perturbative approach at high densities is doubtful, it indicates the possibility of a stabilization mechanism in $f(R)$ gravity. This mechanism, as can be seen from a rapid inspection of Figs. 1 - 4,  leads  to the existence of stable neutron stars {(the stability means that $dM/d\rho_{c}$)} which are more compact objects than in General Relativity. In principle, the observation of such objects could be an experimental probe for $f(R)$ gravity.

\textbf{Model 2}. Let us consider now  the model of quadratic gravity with logarithmic  corrections in curvature \cite{OdintsovLOG}:
\begin{equation}\label{LOG}
f(R)=R+\alpha R^2(1+\beta \ln (R/\mu^2)),
\end{equation}
where $|\alpha|<1$ (in units $r_{g}^{2}$) and  the dimensionless parameter $|\beta|<1$. This model is considered in \cite{Alavirad}  for SLy equation. However it is not  valid beyond the point $R=0$ and we cannot apply our analysis for stars with central density for example $\rho_{c}>1.72\times 10^{15}$ g/cm$^{3}$ (for SLy equation) and at $\rho_{c}>2.35\times 10^{15}$ g/cm$^{3}$ (for FPS equation). The similar situation take place for another. The maximal mass of neutron star at various values $\alpha$ and $\beta$ is close to the corresponding one  in General Relativity at these critical densities (for FPS - $1.75M_{\odot}$, for SLy - $1.93M_{\odot}$). On the other hand,  for model with quark core,  the condition $R=8\pi(\rho^{(0)}-3p^{(0)})>0$ is satisfied at arbitrary densities. The analysis shows that maximal mass is decreasing with growing $\alpha$. By using a piecewise EoS (FPS+quark core) one can obtain  stars with radii $\sim 9.5 $ km and masses $\sim 1.50M_{\odot}$. In contrast with General Relativity,  the minimal radius of neutron star for this equation is $9.9$ km. Using the piecewise (softer) EoS decreases the upper limit of neutron star mass in comparison with using only one EoS for matter within star. For various EoS,  this limit is smaller than $\sim 2M_{\odot}$. Therefore the model with logarithmic corrections does not  lead to new effects in comparison with GR.

\textbf{Model 3}. It is interesting to investigate also the $R^2$ model with a cubic correction:
\begin{equation}\label{CUB}
f(R)=R+\alpha R^{2}(1+\gamma R)\,.
\end{equation}
The case where $|\gamma R| \sim {\mathcal O}(1)$ for  large $R$ is more interesting. In this case the cubic term comparable with quadratic term. Of course we consider the case when $ \alpha R^{2}(1+\gamma R)<<R$. In this case the perturbative approach is valid although the cubic term can exceed the value of quadratic term.
For small masses, the results  coincides with $R^2$ model. For narrow region of high densities, we have the following situation: the mass of neutron star is close to the analogue mass in General Relativity  with  $dM/d\rho_{c}>0$. This means that  this configuration is stable. For $\gamma=-10$ (in units $r_{g}^{2}$) the maximal mass of neutron star at high densities $\rho>3.7\times 10^{15}$ g/cm$^{3}$ is nearly $1.88M_{\odot}$ and radius is about $\sim 9$ km (SLy equation). For $\gamma=-20$ the maximal mass is  $1.94M_{\odot}$ and radius is about $\sim 9.2$ km (see Figs. 5 - 8). In the  General Relativity,  for SLy equation,  the minimal radius of neutron stars is nearly 10 km. Therefore such a model of $f(R)$ gravity can give rise to neutron stars with smaller radii than in General Relativity. Therefore such theory can describe (assuming only the  SLy equation), the existence of peculiar neutron stars with mass $\sim 2M_{\odot}$ (the measured mass of PSR J1614-2230 \cite{Ozel}) and compact stars ($R\sim 9$ km)  with masses $M\sim 1.6-1.7M_{\odot}$ (see \cite{Ozel-2,Guver,Guver-2}).

We also investigate the cases of BSK20 (Figs. 9 - 12) and AP4 EoS (Figs. 13 - 16). The results are similar to the case of SLy EoS: in the context  of gravity with cubic corrections, the existence of more compact neutron stars (in comparison with GR) at high central densities is possible. At the same time, the maximal neutron star mass satisfies the  observational constraints. The parameters of these stable configurations and maximal neutron star mass in model (\ref{CUB}) for these EoS are given in Table I.

Some considerations are needed about the validity of the perturbative approach.  For example, for $\gamma=-20$, $\alpha=10\times 10^{9}$ cm$^{2}$, the relation $\Delta=|\alpha R^{2}(1+\gamma R)/R|$ reaches the maximal value $~0.15$ only at center of star (for the three above mentioned EoS) and only for maximal densities $\rho_{c}$ at which the stable configurations can exist. For more conservative $\gamma=-10$, $\alpha=5\times 10^{9}$ cm$^{2}$, it is  $\Delta_{max}\thickapprox 0.1$.  This allows to apply the perturbative approach with good precision.

{For smaller values of $\gamma$ the minimal neutron star mass (and minimal central density at which stable stars exist) on second branch of stability decreases.}

\begin{table}
\begin{center}
\begin{tabular}{|c|c|c|c|c|c|c|}
\hline
EoS & $M/M_{\odot}$ & $R$, km & $\rho_{c15}$ & $M_{max}/M_{\odot}$ & $\rho_{cmax15}$\\
\hline
SLy & $1.75<M<1.88$ & $9.1<R<9.5$ & $3.34<\rho_{c15}<3.91$ & 2.05 & 2.86 \\
BSK20 & $1.83<M<2.0$ & $9.3<R<9.8$ & $2.97<\rho_{c15}<3.48$ & 2.16 & 2.64 \\
AP4 & $1.95<M<2.07$ & $9.4<R<9.9$ & $2.64<\rho_{c15}<3.09$ & 2.20 & 2.75 \\
\hline
\end{tabular}\end{center}
\caption{The parameters of stability star configurations at high central densities (second branch of stability) for $\gamma=-10$, $\alpha=5\times 10^{9}$ cm$^{2}$ ($\rho_{15}\equiv \rho/10^{15}$g/cm$^{3}$). The maximal mass of neutron star in model (\ref{CUB}) for various EoS is given also. The maximal central density for stability star configurations in General Relativity for corresponding EoS is given in last column for comparison.}
\end{table}

In conclusion we consider the case of cubic modified gravity ($f(R)\approx R+\epsilon R^{3}$). It is interesting to note that for negative and sufficiently large values of $\epsilon$, the maximal limit of neutron star mass can exceed the limit in General Relativity for given EoS (the stable stars exist for higher central densities). Therefore some EoS which ruled out by observational constraints in General Relativity can describes real star configurations in frames of such model of gravity. For example, for BSK19 EoS, the maximal neutron star mass is around $1.86 M_{\odot}$ in General Relativity, but for negative values $|\epsilon|\sim 8$ the maximal mass is around $\sim2M_{\odot}$ (see Fig. 17). The same result holds in the case of FPS, WFF3 and AP2 EoS (at $|\epsilon|\sim 10-12$). For these values of $\epsilon$, these models describe the observational data by \cite{Ozel} with good precision and provide the acceptable upper limit of neutron star mass. One has to note that the  upper limit in this model of gravity is achieved for  smaller radii than in General Relativity for acceptable EoS. Therefore the possible measurement of  neutron star radii with $M\sim 2M_{\odot}$ can give evidences supporting  (or not)  this model of gravity.

\section{Perspectives and Conclusions}

In this paper, we have studied neutron star structures in some $f(R)$ gravity models assuming realistic equations of state.

In particular, we have considered the mass-radius relations for neutron stars  in a gravity models of the form $f(R)=R+R(e^{-R/R_{0}}-1)$ and for the $R^2$ models with logarithmic and cubic corrections. We also investigated the dependence of the maximal mass from the central density of the structure. In the  case of quadratic gravity with cubic corrections,  we found that,  for high central densities ($\rho>10\rho_{ns}$, where $\rho_{ns}=2.7\times 10^{14}$ g/cm$^{3}$ is the nuclear saturation density)  stable star configurations  exist. In other words, we have a  second ``branch'' of stability with respect to the one existing in General Relativity. The minimal radius of such stars is close to $9$ km with maximal mass
$\sim 1.9 M_{\odot}$ (SLy equation), $\sim 2.0 M_{\odot}$ (BSK20 equation), $\sim 2.07 M_{\odot}$ (AP4 equation). This effect  gives rise to more compact stars than in General Relativity and could be extremely relevant  from an  observational point of view. In fact,  it is interesting to note that using an equation of state in the framework  of $f(R)$ gravity with cubic term gives rise to two important features: the existence of an upper limit on neutron star mass ($\sim2M_{\odot}$) and the existence of neutron stars with radii $R\sim 9 \div 9.5$ km and masses $\sim 1.7M_{\odot}$. These facts could have a twofold interest: from one side, the approach could be useful to explain  peculiar objects that evade  explanation in the framework of standard General Relativity (e.g. the magnetars \cite{magnetar}) and, from the other side, it could constitute a very relevant test for alternative gravities.  Another interesting result can be realized in cubic modified gravity. Some EoS, ruled out   in General Relativity according  to observational data,  satisfy the observational constraints in this model and give a  realistic description of $M$-$R$ relation and acceptable upper limit of neutron star mass.

\acknowledgments

A.V.A. would like to thank A.V. Yaparova for useful discussion {and Prof. K.Y. Ek\c{s}i for providing the observational constraints data}. S.C. and S.D.O. acknowledge the support of the visitor program  of the Kobayashi-Maskawa Institute for the Origin of Particles and the Universe (Nagoya, Japan). S.C. acknowledges the support of INFN (iniziativa specifica TEONGRAV). S.D.O. acknowledges the support of visitor program of Baltic Federal University and useful discussions with Prof. A. Yurov. This work has been supported in part by MINECO (Spain), FIS2010-15640, AGAUR (Generalitat de Catalunya), contract 2009SGR-345, and MES project 2.1839.2011 (Russia) (S.D.O.).

\begin{figure}
  \includegraphics[scale=1.1]{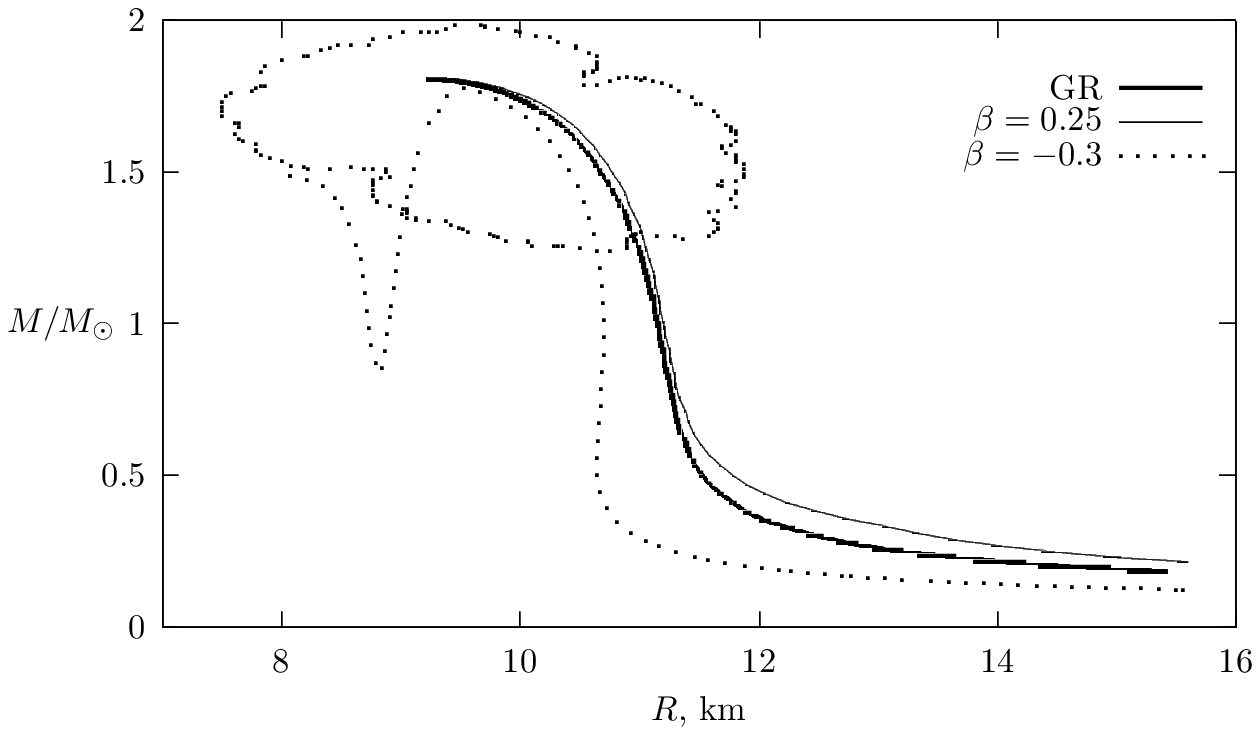}\\
  \caption{The mass-radius diagram for neutron stars in $f(R)$ model (\ref{EXP}) in comparison with General Relativity by using a  FPS equation of state. {The constraints derived from observations of three neutron stars from \cite{Ozel} is depicted by the dotted contour (hereinafter). For negative values of $\beta$ for high central densities we have the possibility of existence of stable star configurations ($dM/d\rho_{c}>0$). If $\beta=-0.3$ the masses of these configurations are $0.85M_{\odot}<M<1.48M_{\odot}$ ($8.1<R<8.8$ km)}}.
\end{figure}

\begin{figure}
  \includegraphics[scale=1.1]{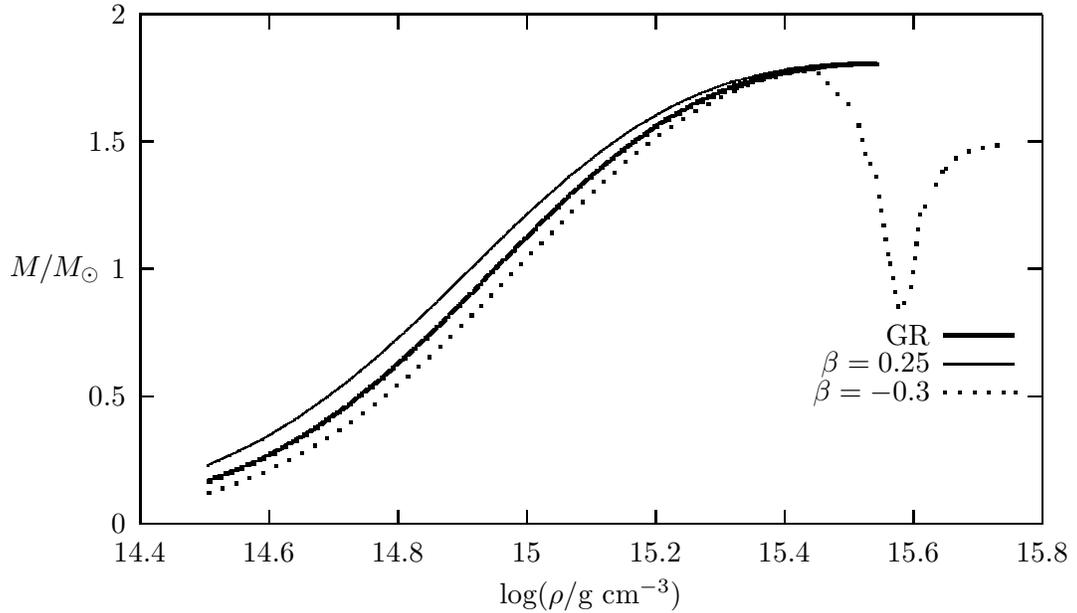}\\
  \caption{The dependence of neutron star mass from central density in $f(R)$ model (\ref{EXP}) in comparison with General Relativity  for a  FPS equation of state. In narrow interval of central densities $3.76\times 10^{15}<\rho_{c}<5.25\times 10^{15}$ g/cm$^{-3}$ the stability star configurations ($dM/d\rho_{c}>0$) exist. For comparison the upper limit of central density in GR is $3.48\times10^{15}$ g/cm$^{3}$ for FPS EoS.}
\end{figure}

\begin{figure}
  \includegraphics[scale=1.1]{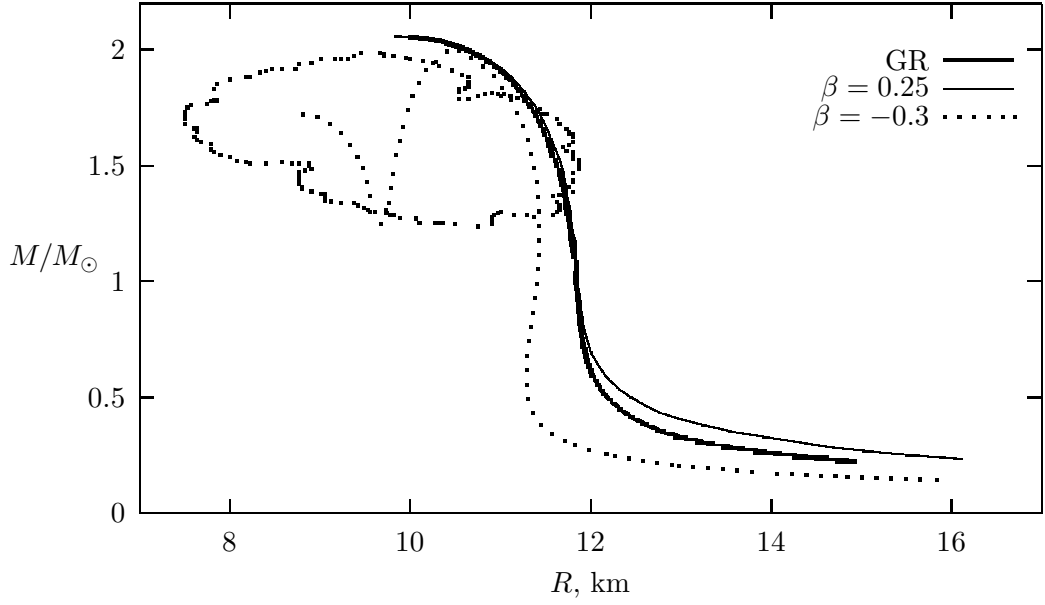}\\
  \caption{The mass-radius diagram for neutron stars in $f(R)$ model (\ref{EXP}) in comparison with General Relativity for a SLy equation of state. {For $\beta=-0.3$ in this model there are stable stars with masses $1.25_{\odot}<M<1.72M_{\odot}$ and radii $8.8<R<9.7$ km. Therefore the description of observational constraints is more better than in GR for SLy EoS.}}
\end{figure}

\begin{figure}
  \includegraphics[scale=1.1]{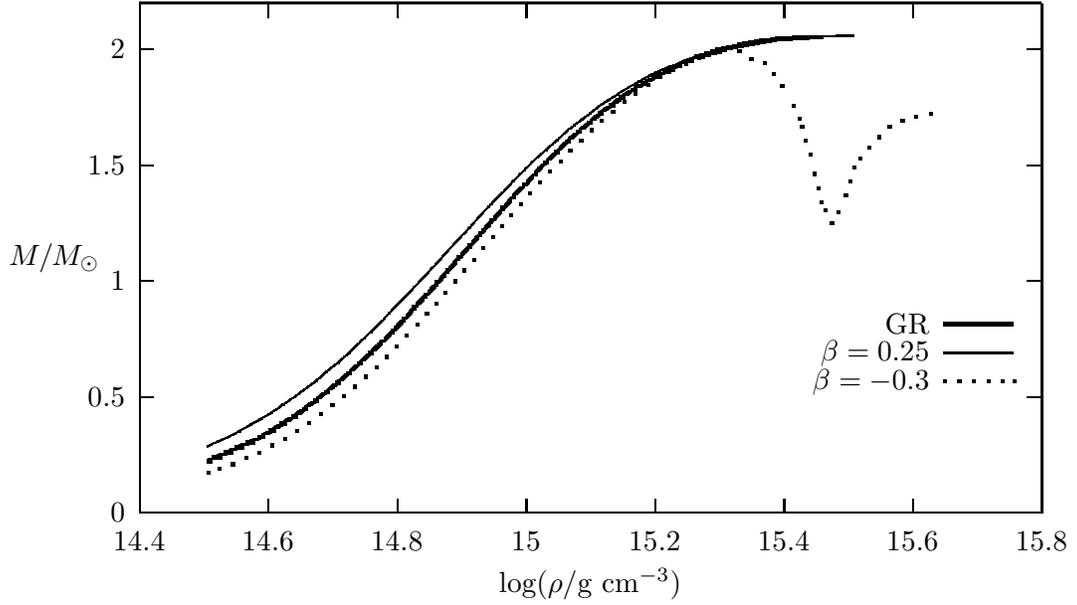}\\
  \caption{The dependence of neutron star mass from central density in $f(R)$ model (\ref{EXP}) in comparison with General Relativity for Sly equation of state. {The stable stars don't exist in GR for $\rho_{c}>2.86\times 10^{15}$ g/cm$^{3}$ while for model (\ref{EXP}) with $\beta=-0.3$ such possibility can take place at $2.97\times10^{15}<\rho_{c}<4.23\times10^{15}$ g/cm$^{3}$.}}
\end{figure}

\begin{figure}
  \includegraphics[scale=1.1]{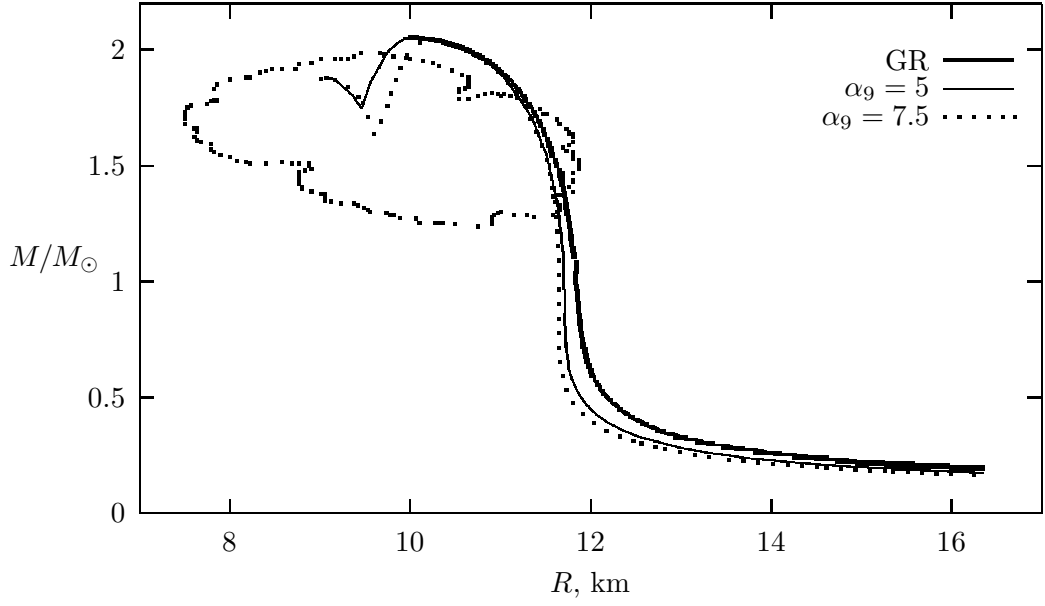}\\
  \caption{The mass-radius diagram for neutron stars in $f(R)$ model with cubic corrections (\ref{CUB})($\gamma=-10$) in comparison with General Relativity assuming a SLy EoS. The notation $\alpha_{9}$ means $\alpha_{9}=\alpha/10^{9}$ cm$^{2}$.}
\end{figure}

\begin{figure}
  \includegraphics[scale=1.1]{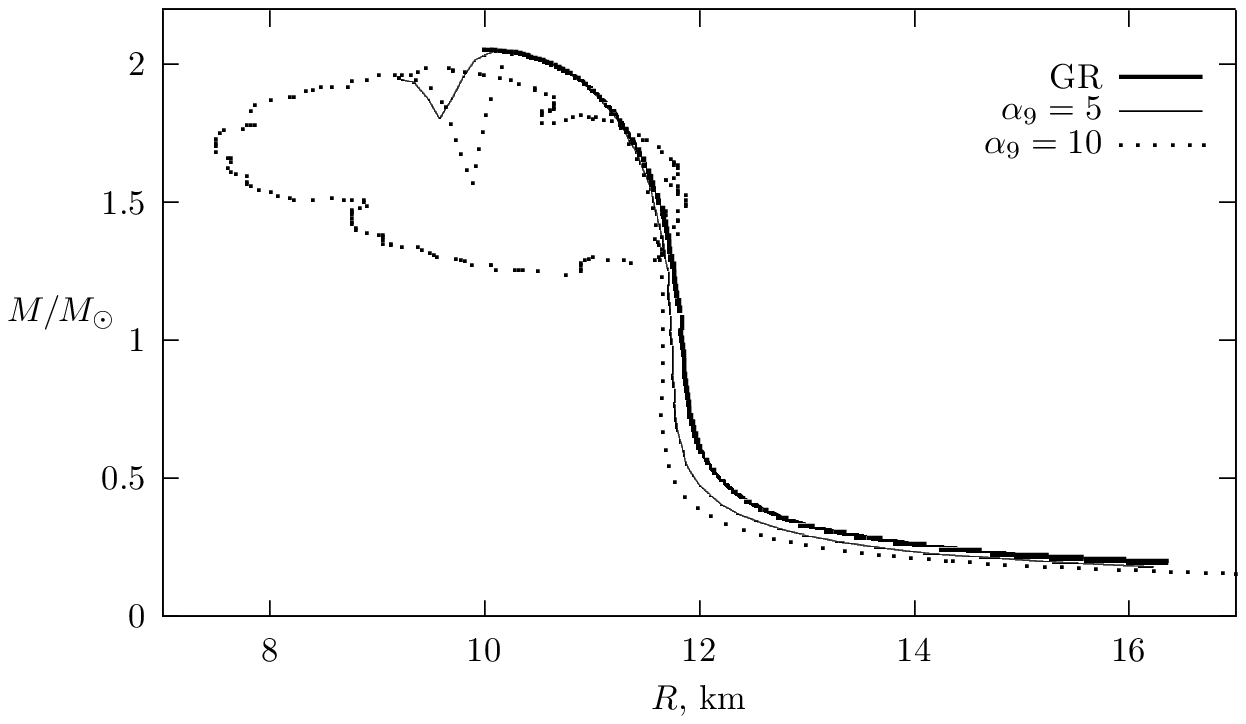}\\
  \caption{The dependence of neutron star mass from central density in $f(R)$ model (\ref{CUB}) ($\gamma=-10$) for SLy EoS.}
\end{figure}

\begin{figure}
  \includegraphics[scale=1.1]{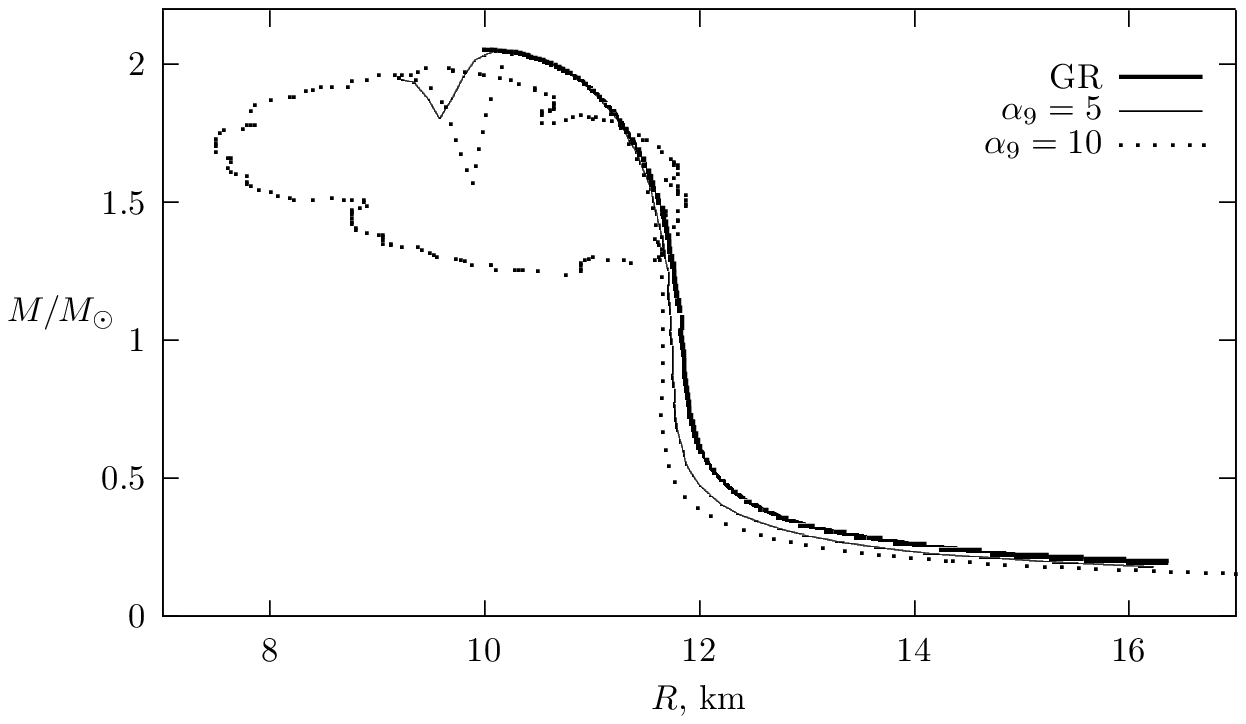}\\
  \caption{The mass-radius diagram for neutron stars in $f(R)$ model with cubic corrections (\ref{CUB})($\gamma=-20$) for SLy EoS.}
\end{figure}

\begin{figure}
  \includegraphics[scale=1.1]{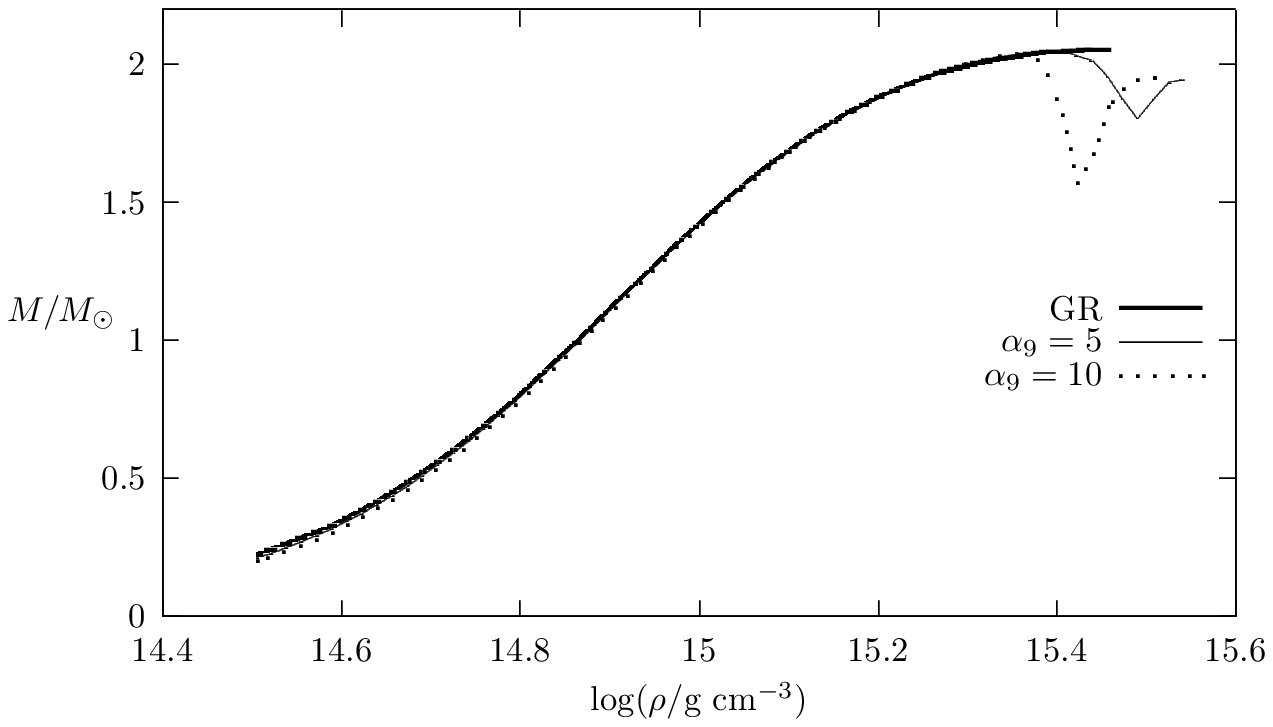}\\
  \caption{The dependence of neutron star mass from central density in $f(R)$ model (\ref{CUB}) ($\gamma=-20$) for SLy EoS.}
\end{figure}

\begin{figure}
  \includegraphics[scale=1.1]{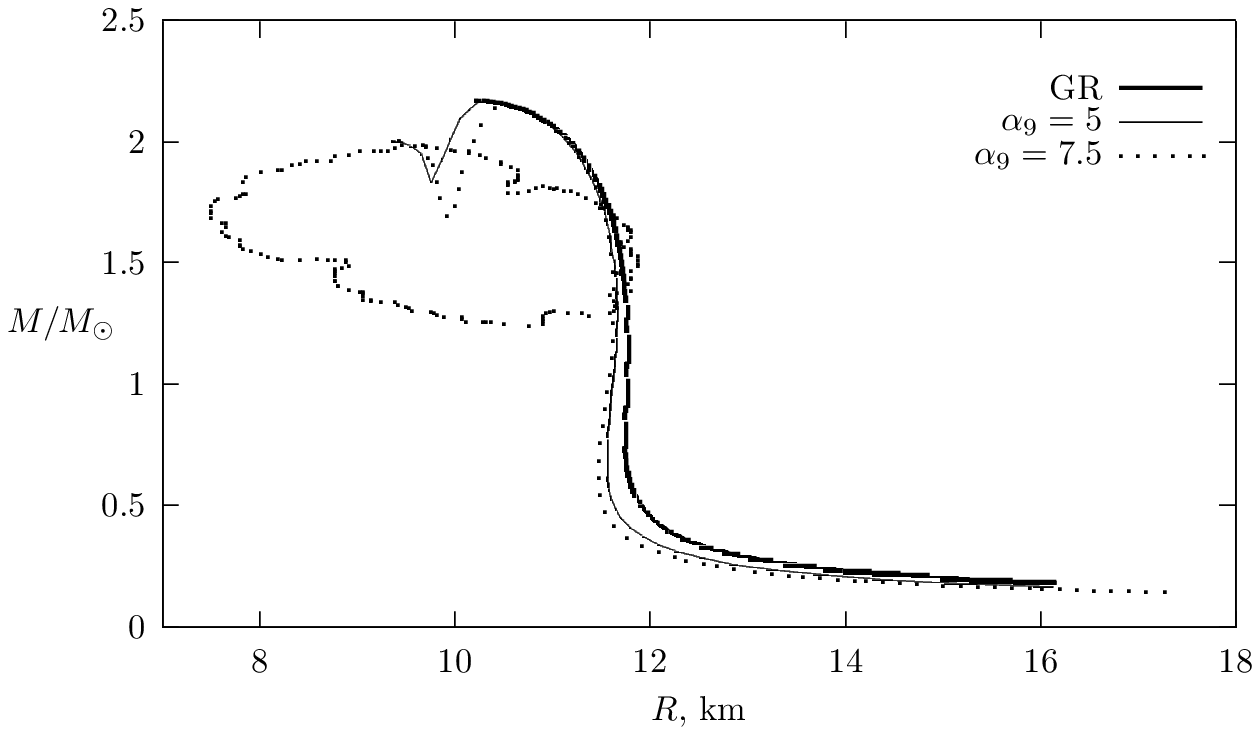}\\
  \caption{The mass-radius diagram for neutron stars in $f(R)$ model with cubic corrections (\ref{CUB})($\gamma=-10$) for BSK20 EoS.}
\end{figure}

\begin{figure}
  \includegraphics[scale=1.1]{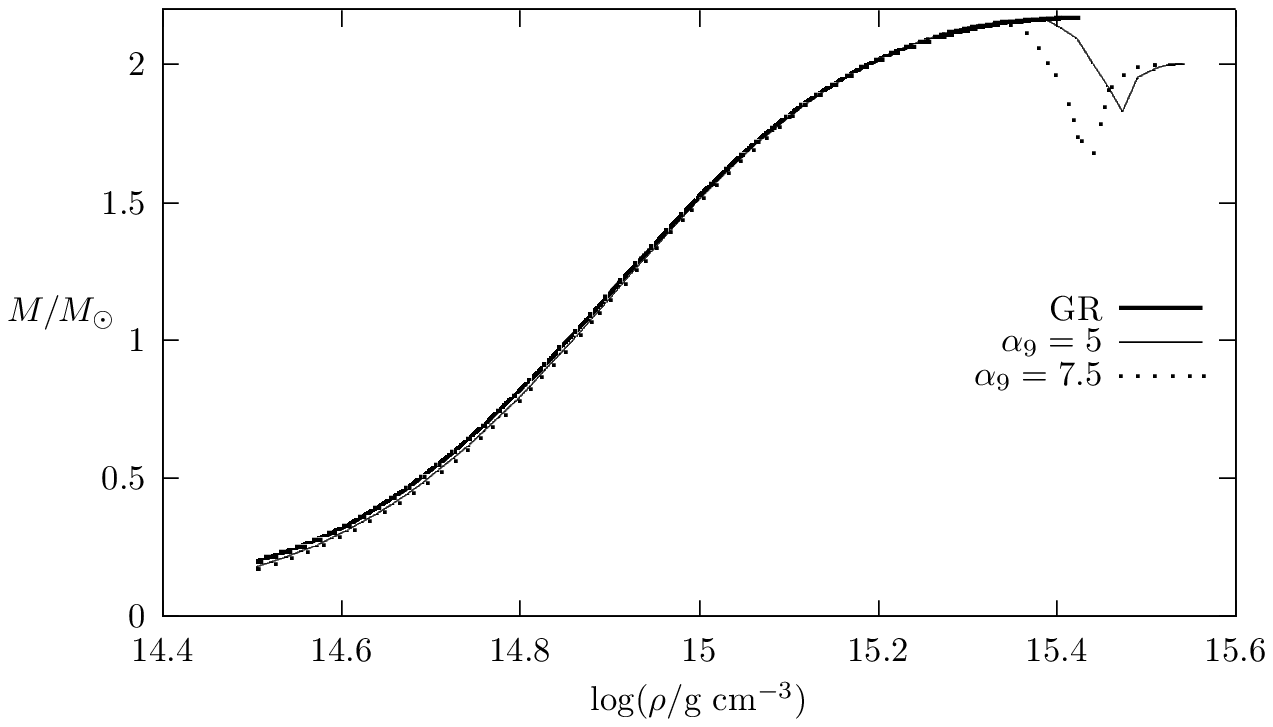}\\
  \caption{The dependence of neutron star mass from central density in $f(R)$ model (\ref{CUB}) ($\gamma=-10$) for BSK20 EoS.}
\end{figure}

\begin{figure}
  \includegraphics[scale=1.1]{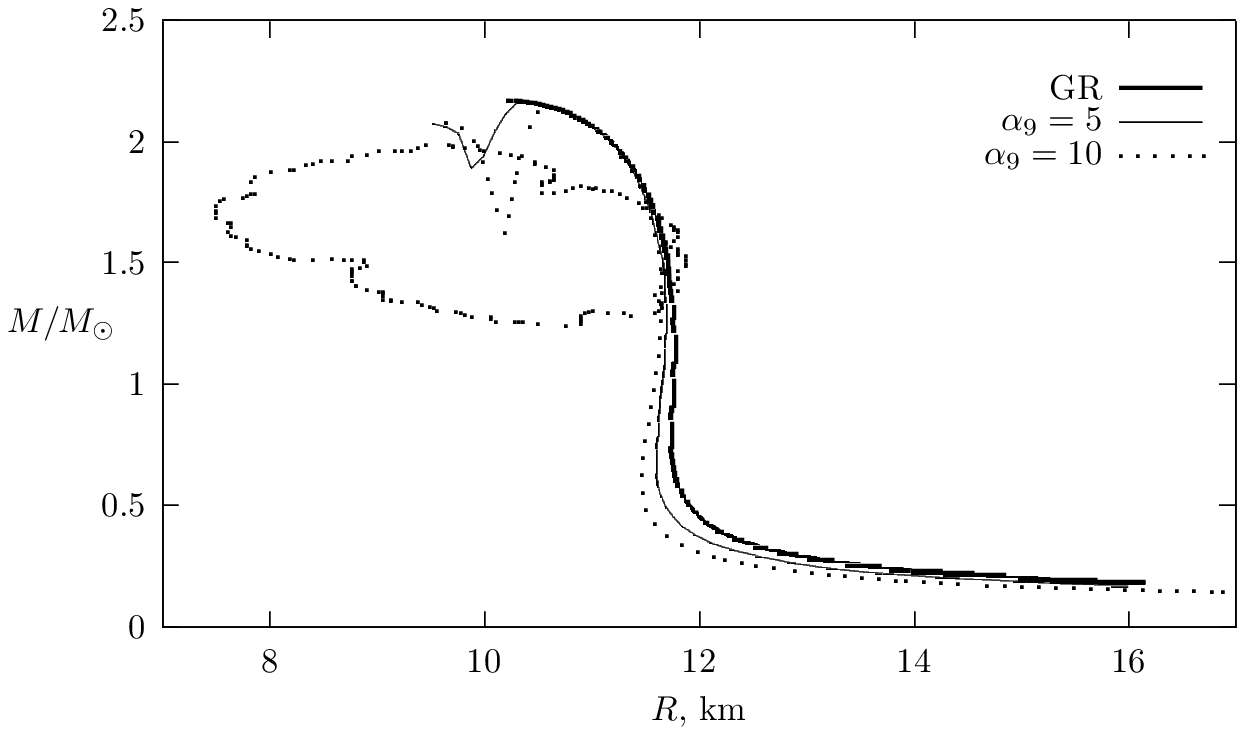}\\
  \caption{The mass-radius diagram for neutron stars in $f(R)$ model with cubic corrections (\ref{CUB})($\gamma=-20$) for BSK20 EoS.}
\end{figure}

\begin{figure}
  \includegraphics[scale=1.1]{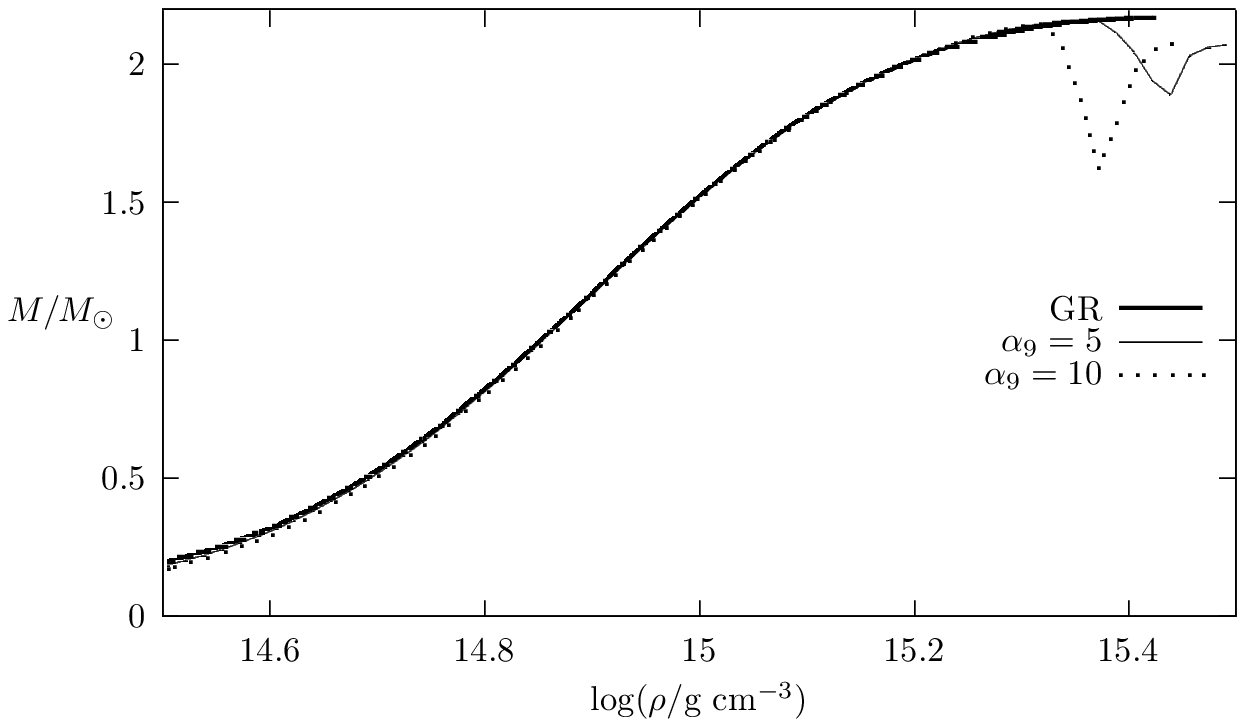}\\
  \caption{The dependence of neutron star mass from central density in $f(R)$ model (\ref{CUB}) ($\gamma=-20$) for BSK20 EoS.}
\end{figure}

\begin{figure}
  \includegraphics[scale=1.1]{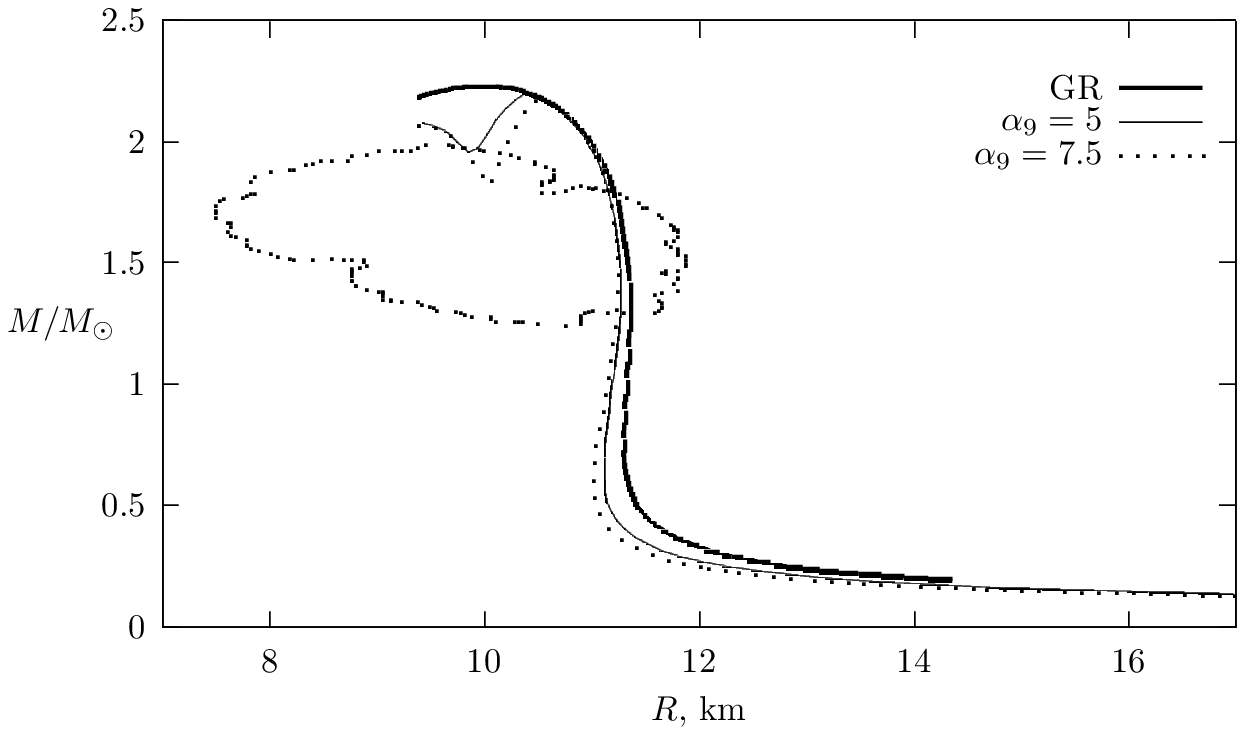}\\
  \caption{The mass-radius diagram for neutron stars in $f(R)$ model with cubic corrections (\ref{CUB})($\gamma=-10$) for AP4 EoS.}
\end{figure}

\begin{figure}
  \includegraphics[scale=1.1]{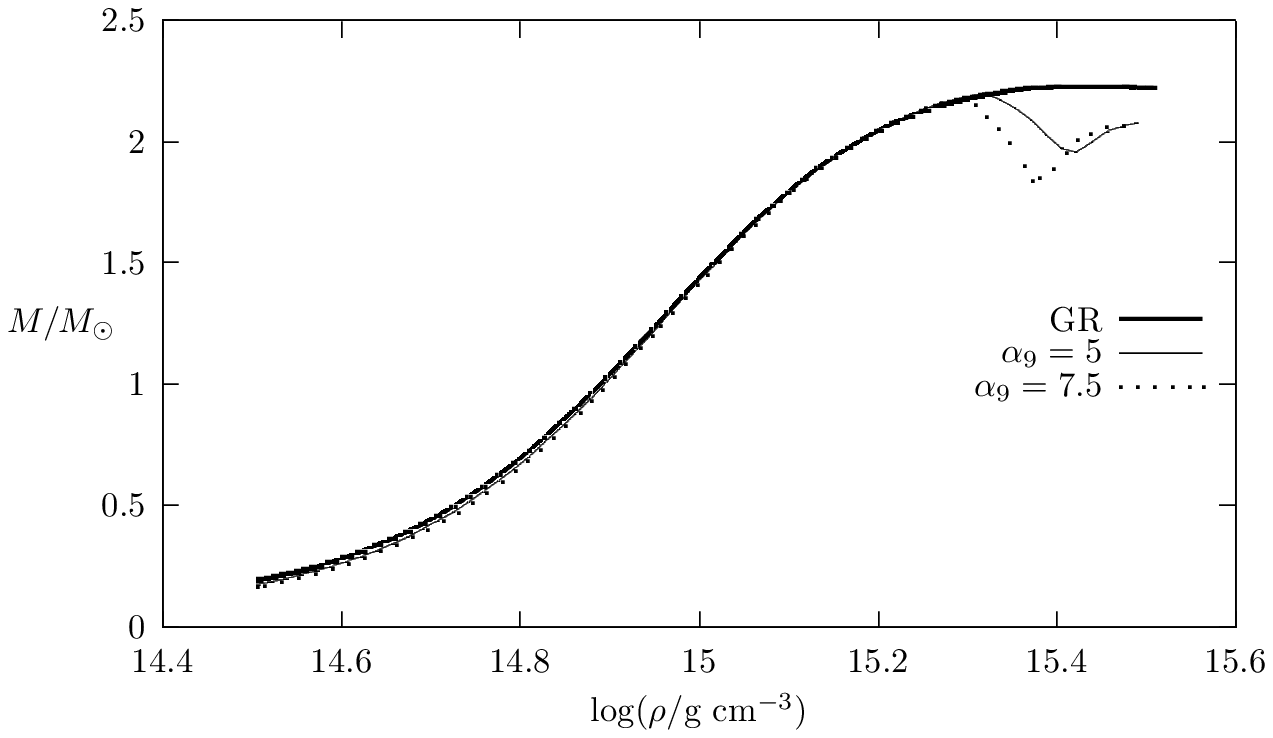}\\
  \caption{The dependence of neutron star mass from central density in $f(R)$ model (\ref{CUB}) ($\gamma=-10$) for AP4 EoS.}
\end{figure}

\begin{figure}
  \includegraphics[scale=1.1]{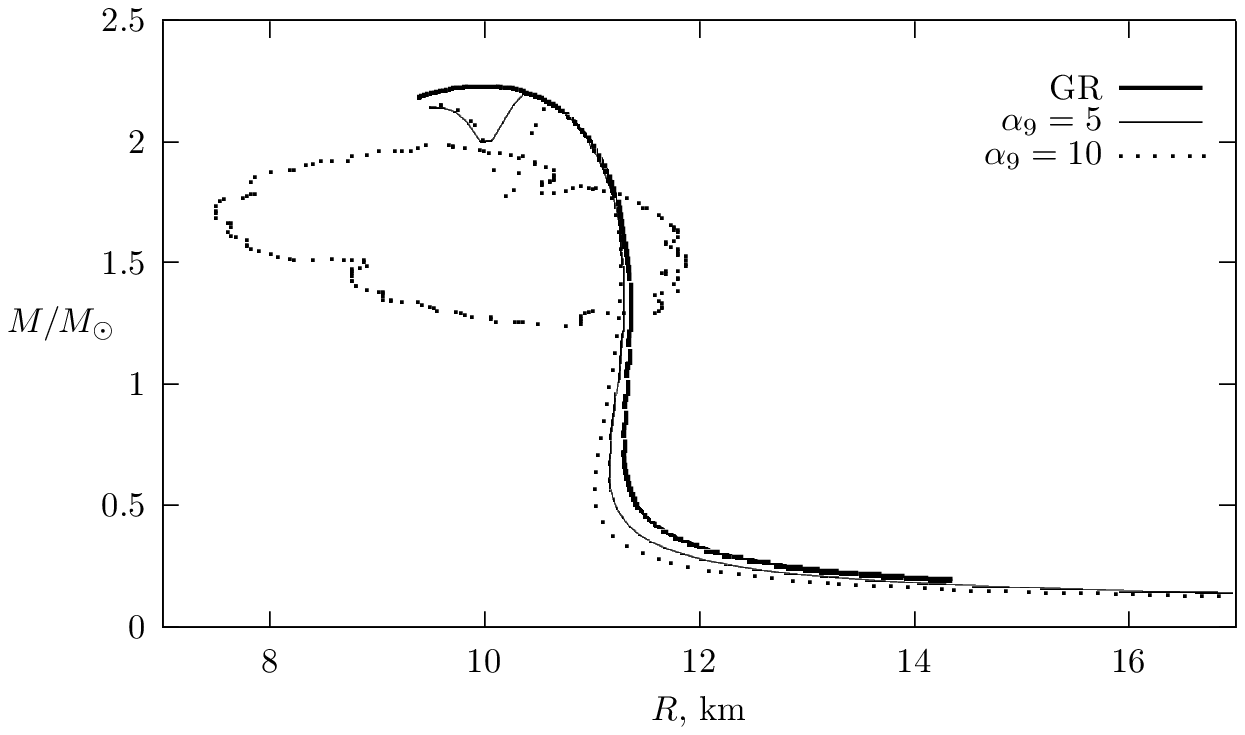}\\
  \caption{The mass-radius diagram for neutron stars in $f(R)$ model with cubic corrections (\ref{CUB})($\gamma=-20$) for AP4 EoS.}
\end{figure}

\begin{figure}
  \includegraphics[scale=1.1]{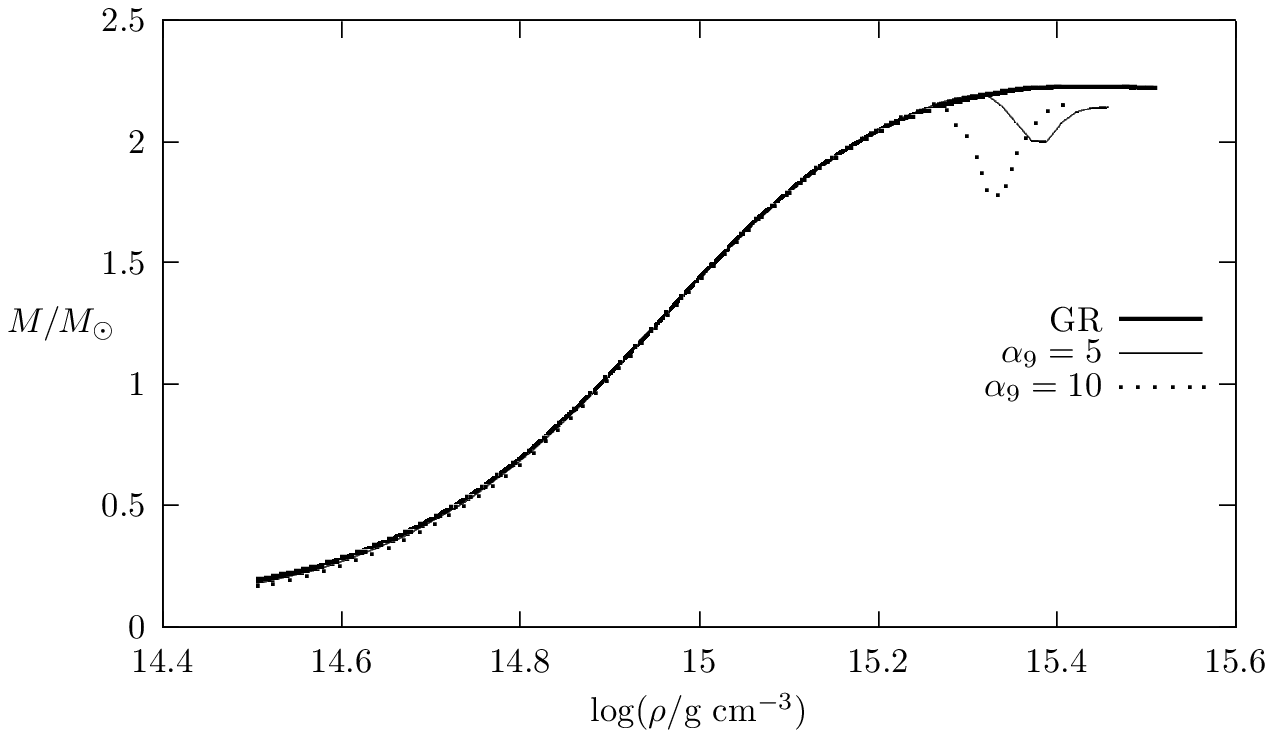}\\
  \caption{The dependence of neutron star mass from central density in $f(R)$ model (\ref{CUB}) ($\gamma=-20$) for AP4 EoS.}
\end{figure}

\begin{figure}
  \includegraphics[scale=1.1]{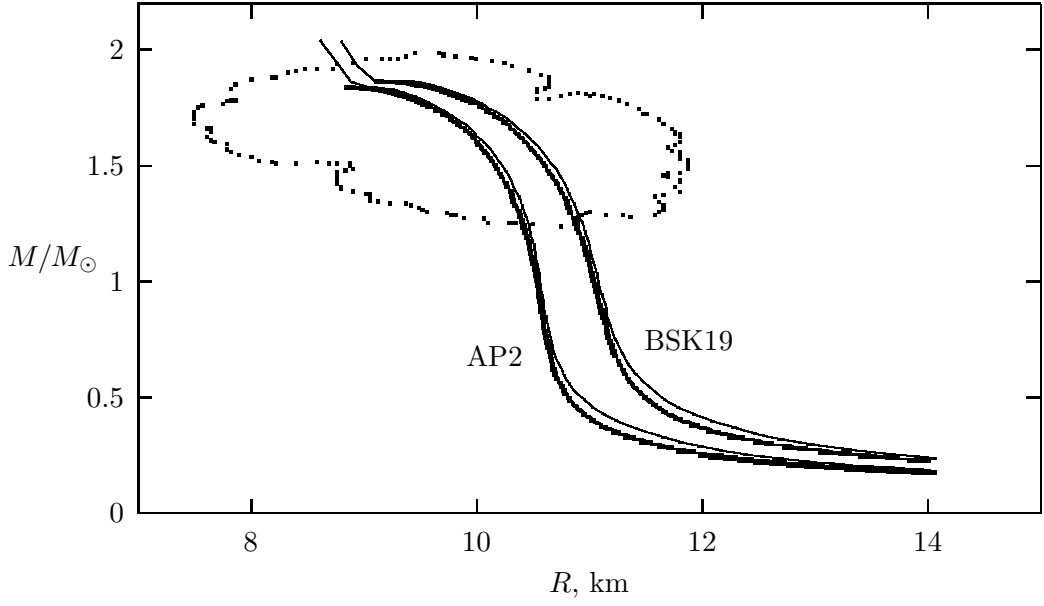}\\
  \caption{The mass-radius diagram for neutron stars in $f(R)\approx R+\epsilon R^{3}$ model for AP2 and BSK19 EoS (thick lines) in comparison with General Relativity (bold lines). The parameter $\epsilon=-10$ (in units of $r_{g}^{4}$. In frames of model of modified gravity these EoS give the upper limit of neutron star mass $\sim 2M_{\odot}$ (the corresponding radius is $R~8.6-8.8$ km and describe observational data from \cite{Ozel} with acceptable precision. With increasing $|\epsilon|$ the upper limit of mass increases. The similar effect one can seen for WFF3 and FPS EoS at slightly larger $|\epsilon|$.}
\end{figure}

\end{document}